# Geometric Meditation-Based Cognitive Behavioral Therapy in Obsessive-Compulsive Disorder: A Case Study

Fatemeh Torkamani[1,2,3,4], Farshad Nazaraghaie[2,3,4], Mohammad Nami[1,4,5,6]*

[1]DANA Brain Health Institute, Iranian Neuroscience Society, Fars Chapter, Shiraz, Iran
[2]Fars Meditation Academy, Shiraz, Iran
[3]Social and Individual Stress Reduction Association, Shiraz, Iran
[4]Neuroscience Laboratory (Brain, Cognition and Behavior), Department of Neuroscience, School of Advanced Medical Sciences and Technologies, Shiraz University of Medical Sciences, Shiraz, Iran
[5]Department of Neuroscience, School of Advanced Medical Sciences and Technologies, Shiraz University of Medical Sciences, Shiraz, Iran
[6]Academy of Health, Senses Cultural, Sacramento, CA, USA

*Corresponding author:
Mohammad Nami MD, PhD, Department of Neuroscience, School of Advanced Medical Sciences and Technologies, Shiraz University of Medical Sciences, Shiraz, Iran. torabinami@sums.ac.ir

*Abstract*

Obsessive-Compulsive Disorder (OCD), characterized by unwanted and distressing intrusive thoughts, images, urges, doubts and ideas or sensations and repetitive mental or behavioral acts, which is regarded as an overwhelming mental disorder. Over the past few years, several studies have indicated how mindfulness-based interventions may be effectively used to remediate OCD symptoms based on which such methods are endorsed as effective complementary or alternative options to remediate OCD. The present pilot investigation assessed the effectiveness of Geometric Meditation-based Cognitive Behavior Therapy (GM-CBT) as a novel integrated approach to help OCD symptoms. Accordingly, an eight-week treatment program (90 minute sessions per week) in a single case of intractable OCD was found to result in a significant reduction in OCD symptoms, anxiety and depression as well as increased mindfulness skills and subsequent secondary outcomes. A three-month post treatment follow up suggested long-lasting beneficial effects. Such a pilot model may receive further endorsement as a holistic CBT approach for OCD.

## 1. Introduction

Obsessive-Compulsive Disorder (OCD) is characterized by unwanted and distressing intrusive thoughts, images, urges, doubts and ideas or sensations as well as repetitive mental or behavioral acts performed in response to an obsession to prevent or reduce anxiety, distress, or unwanted events. The patient may either have enough insight about his disease or minimal to no insight at all (1). There are two types of resistance strategies to obsessions. Those include overt compulsive rituals through repetitive behaviors (washing, cleaning, checking) and covert compulsive rituals to suppress the unwanted thoughts, neutralization and avoidance (2). Many OCD sufferers have dysfunctional beliefs such as catastrophizing, perfectionism, excessive responsibility, the need to control thoughts, overvalued thoughts and the inability to tolerate ambiguity and uncertainty. Such false beliefs which potentially lead to dysfunctional strategies are referred to as maintaining factors in OCD (1).

OCD is the fourth leading psychiatric morbidity with its prevalence and annual incidence ranging from 1-3% and 1-2% in general population, respectively (3). Along these lines, anxiety-related disorders are known as key comorbid conditions in OCD (4). Based on some reports, almost 15% of people with OCD fulfilled criteria for generalized anxiety disorder, while 20, 36, 30 and 54% were found to have comorbid panic disorder, social phobia, specific phobias, and relapsed major depressive disorder, respectively (4). In addition, some studies have suggested that OCD may comorbid with Obsessive-Compulsive Personality Disorder (OCPD) classified as a specific subtype of OCD (5-6).

With regard to treatment, the traditional cognitive behavioral therapy (CBT) through exposure and response prevention (ERP) and/or drug treatments with selective serotonin reuptake inhibitors (SSRIs) are considered as common approaches to treat OCD (7-9). Meanwhile, up to one third of patients are found be refractory to ERP-based CBT or medication therapy (10-11).

For the first time, the concept of meditation inspired by Zen Buddhism was introduced by Fritz Perls (1969) in the form of "increased awareness in the present moment" through Gestalt therapy. The concept found its way into clinical treatment in the 1970s under the term "mindfulness" (12). The combination and integration of the concepts of "acceptance" and "mindfulness" with traditional CBT led to a so-called "third wave" of cognitive-behavioral therapies (13-14).

Based on recent investigations, mindfulness-based cognitive therapy (MBCT) has been a successful approach for the treatment of OCD patients who have not responded to conventional treatments and/or in reducing the residual symptoms of OCD disorder (15-17). The beneficial impact of MBCT in OCD is potentially based on remediating cognitive predicaments mainly in attentional functions, emotion regulation and executive functions (15). A new study reported the beneficial effects of MBCT in OCD, depression and anxiety symptoms as well as obsessive beliefs through improving mindfulness skills and self-compassion in the experimental group vs. controls (18). In addition, other third wave treatment approaches such as Acceptance and Commitment Therapy (ACT) and Mindfulness Based Stress Reduction are also reported as beneficial in relapsed OCD symptoms (14, 19-21).

Farshad's Geometric Meditation (in short, F.G. Meditation) is a novel holistic style of meditation, based on the principles of mathematics, geometry, psychology, physiology, neurosciences, cognitive sciences and philosophy of mind comprising 64 techniques in three fields of concentrative, introspective and mindfulness. They are generally classified in six categories as follows: Geometric Somatic-Breathing Based Meditation (GSBBM), Geometric Introspective Based Meditation (GIBM), Mindfulness-Based Meditation (MBM), Somatic Based Meditation (SBM), Somatic-Breathing Based Meditation (SBBM) and Mind-Body Based Meditation (MBBM) (22). These techniques tend to stimulate various domains of awareness including sensational-physical awareness, emotional-mental awareness and existential awareness (23). F.G. Meditation is a self-induced and intentional manipulation of attention leading to an enhanced states of consciousness and focused attention (24, 25).

A recent qEEG-based and polygraphic study on F.G. Meditation demonstrated an increased central theta/beta relative power in central cortical brain regions corresponding to deep relaxation state in test condition compared to the control as well as changes in brain activity were identified before, during and after a number of geometric mediation techniques. (26-27). In addition, ten sessions of mindfulness-based F.G. Meditation led to a notably decreased depressive symptoms, anxiety and OCD in middle-aged women and the significant increase was observed in their interpersonal relationships. Results were followed 3 months after the end of treatment (28). Besides, a similar study revealed that three month F.G. Meditation training had a significant relief in clinical symptoms of patients with asthma (29).

While earlier studies have shown the efficacy of Mindfulness-based meditation in OCD, F.G. Meditation techniques have not yet been applied in treatment of OCD. Some key features on F.G. Meditation including its objective-rational perspective yet a high emotional potential; lack of ideological bias; variation in techniques; easy training; extended states of physical-emotional relaxation; helping to dampen emotional thinking; finding other dimensions of processes and contents of mind; having an educational animation and audio files to prepare for practice at home make it worthwhile to examine its role in possibly remediating symptoms in mental illnesses including OCD.

That said, the aim of this report was to describe a pilot study in which a person with OCD participated in tailored F.G. Meditation-Based Cognitive Behavioral Therapy (GM-CBT).

## 2. Case Presentation

Ms. A.B., 40-year-old single woman who lived in a family of 7 persons, holding an undergraduate degree and working as founding director of a private institute. Her obsession started about 12 years ago with the content of unwanted sexual thoughts and urges. She also reported other obsessions and rituals including the behavior of checking (locks, gas valves, and doors), symmetry and order but none of them were not annoying and unbearable for her except obsessive sexual thoughts and feelings.

Her obsessive disorder was disturbing enough to prevent her from continuing education and professional development activities, and also had a negative effect on interpersonal relationships with her family, colleagues, clients and friends. She was in conflict for 12 years to convince herself to take action for the treatment.

She did not believe in medications to treat her disorder and refrained from visiting her psychiatrist. Upon presentation, her coping strategies for dealing with obsessions were covert compulsions such as the suppression of thoughts, positive self-talking, and positive suggestion, and also overt compulsions like the gaze to the people, prevent blinking, and widen eyes at the time of rising sexual thoughts and feelings. Given the tic-like compulsive behaviors she complained of persistent contractions and pain in the muscles around her eyes. Meanwhile, she denied a past history of a tic disorder.

She was also complaining of inadequate quality of sleep, difficulty with anger management, and social anxiety. She primarily underwent a set of psychological assessments using the tools including: 1- Minnesota Multiphasic Personality Inventory – 2 (MMPI-2) (30, 31); and 2- Millon Clinical Multiaxial Inventory – III (MCMI-III) (32-34); and 3- Yale-Brown Obsessive Compulsive Scale (YBOCS) (35-36).

## 2.1. Evaluation outcome

A: The clinical interview revealed obsession with the symptoms of anxiety, depression and some obsessive-compulsive personality traits, such as perfectionism, double oriented, the need for social approval, and inflexibility, especially in dealing with new situations. Also, she had a history of an unwanted caressing without consent during her adolescence, known as Sibling Abuse (2 times). Other problems were dissatisfaction with the quality of sleep (insufficient sleep), anger management problems, and performance anxiety (limited to speech).

B: For the MMPI-2, all major scales were within normal limits (T scores 50 to 60) except scale 4 (Psychopathic Deviate). Significantly high score in this scale (T=77) indicates family conflict, inconsistent with authorities, and the tendency to anger and conflict with others. Others know them as talkative, extroverted and active individuals but furious and self-centered. Also, MMPI-2 Content Scales showed high scores in anxiety (T=72), obsession (T=75), depression (T=69), and anger (T=70) which was fully in line with the results of interviews. The assessment indicated irritability, resentment, anger management difficulties, as well as possibility to experience physical abuse.

C: Millon Clinical Multiaxial Inventory – III (MCMI-III) results showed a relatively high score (T=74) in scale 7 (compulsive). Base rate scores of 75 or greater indicate that the examiner should proceed with caution. Also, anxiety disorder score (clinical syndrome scale) was 67.

## 2.2. Treatment options

Following evaluations, four treatment options were recommended including: 1-Medication therapy under the care of a psychiatrist; 2-Exposure Response

Prevention; or 3- The combination of the two. The fourth option was Geometric Meditation Based Cognitive Behavioral Therapy (GM-CBT) which was followed by the patient.

The patient signed an informed consent before entering the study. The entire process was in accordance with the ethical standards laid down by the institutional review board at Shiraz University of Medical Sciences.

### 2.3. Assessment tools
### 2.3.1. Tests

Four sets of evaluations were done to examine the effectiveness of our intervention on the patient's obsessive compulsive symptoms at baseline and after 8 sessions.

1. *Minnesota Multiphasic Personality Inventory – 2 (MMPI-2)(567 Q)*
2. *Yale-Brown Obsessive Compulsive Scale (YBOCS)*
3. *Obsessional Beliefs Questionnaire - Revised (OBQ-44)*
4. *Five-Facet Mindfulness Questionnaire*

The OBQ-44 is a 44-item self-report measure of cognitive beliefs and maintenance factors associated with OCD which was developed by the Obsessive Compulsive Cognitions Working Group (OCCWG). The instrument is related to the etiology and maintenance of obsessions and compulsions. It has three subscales: (a) Responsibility / Threat Estimation, (b) Perfectionism / Certainty, and (c) Importance / Control of thoughts. The scales have excellent internal consistency ($\alpha > .89$ for each subscale). OBQ-44 distinguishes between people diagnosed with OCD and non-OCD anxious controls (37). Validity and reliability of Persian form of OBQ - 44 have been studied in the Iranian population. It has been shown that this questionnaire could be used as a valid test (38). The psychometric properties of the Persian version indicated five factors: (a) general; (b) perfectionism / certainty; (c) responsibility and threat estimation; (d) importance and control of thoughts; (e) complete performance (39). The minimum and maximum possible scores are -132 and +132. The more negative score is indicated the high level of obsessive thoughts in person. Conversely, higher positive scores reflect the balance of thoughts in person.

The Five-Facet Mindfulness Questionnaire is a trait mindfulness measure 39-item self-report scale that is made by Baer (2006) et al. (40). This scale was created through a combination of items from other types of mindfulness questionnaire and developed using factor analysis. The five facets are observing, describing, acting with awareness, non-judging of inner experience, and non-reactivity to inner experience. These statements represent on a 5-point Likert scale from 1 = never or very rarely true to 5 = very often or always true (40). Based on Neuser research (2010), the internal consistency was good and alpha coefficient was in a range of 0.75 (non-reactivity factors) to 0.91 (description factors) (41).

This questionnaire has been validated in Iranian population and its questions have been reduced to 31. Factor analysis showed that the questions 31, 27, 25, 24, 19, 14, failed to load on any factor, and so were excluded from the total of 39 items of the questionnaire. Also, the number of factors in this questionnaire increased from five to six i.e. (1) description of inner experience; (2) concentration and mindfulness; (3) non-judging the inner experience; (4) observing thoughts; emotions and physical senses; (5) acting with awareness; and (6) adjust reactions to events. Cronbach's alpha coefficient for the questionnaire was reported 0.79 and for six factors was respectively 0.87, 0.88, 0.82, 0.84, 0.78, and 0.84. In this questionnaire, the highest score that can be obtained is 155, and the lowest score is 31. Closer scores to 155 indicate higher mindfulness and vice versa (42).

### 2.3.2. Close-open ended questions (upon final session)

a) Were the treatment sessions effective for you? Which part of that you found most effective?
b) In what ways this treatment affected your condition?
c) Do you have enough self-discipline to practice the techniques after completing the treatment on a daily basis?
d) What was your favorite technique? And what was more effective for you?
e) Have has meditation exercise been effective in other areas of your life and how?
f) What other areas in your life you would expect to be affected by the practice of meditation?

### 2.3.3. Self-reported experience (upon follow-up)

The qualitative reports of client experiences were recorded during eight sessions and follow-up period on monthly basis (15).

## 3. Study design

The present case investigation examined the effect of F.G. Meditation in a single case of OCD (as per the DSM-5 criteria) following 8 sessions of 90 minute training once a week. Measures were taken at baseline, post-therapy and at 3 months post-therapy (Figure 1).

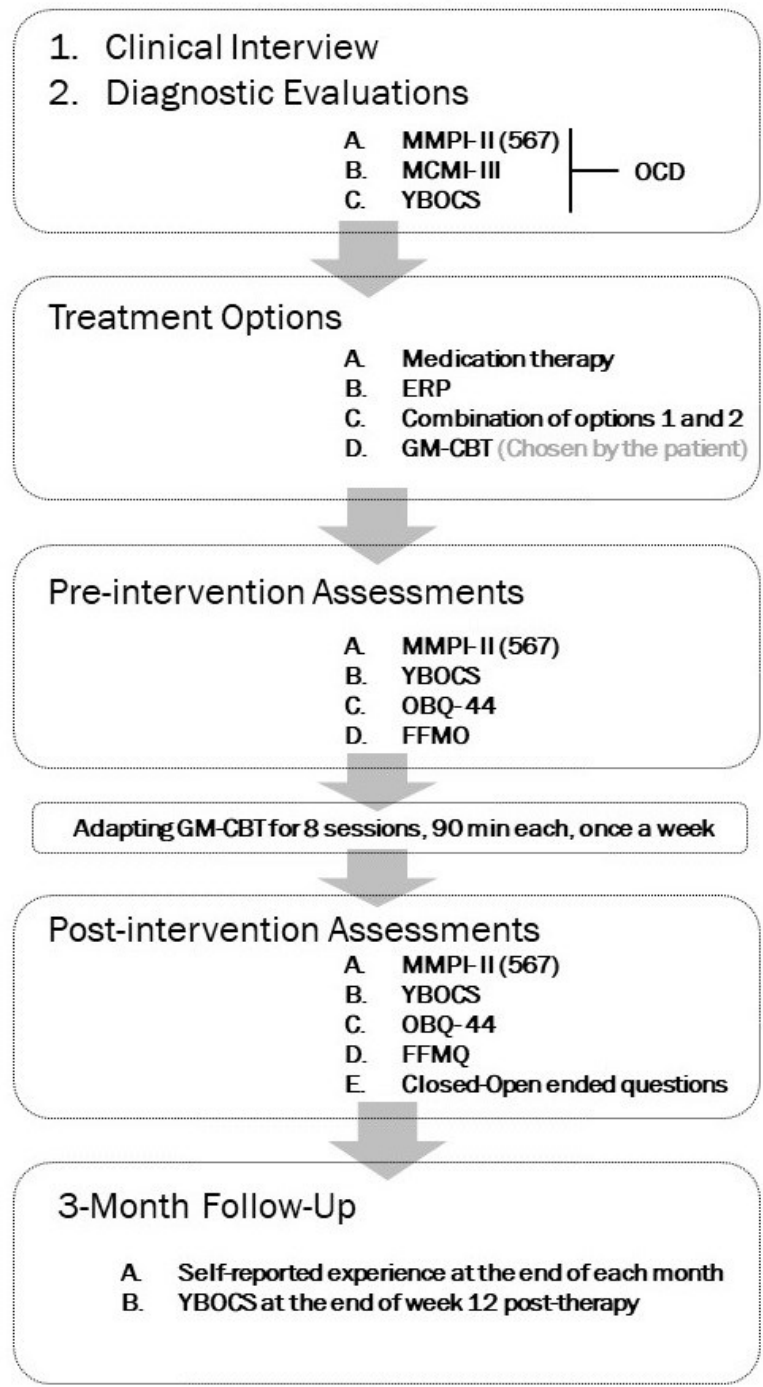

Figure 1. The study protocol

## 4. Intervention

### 4.1. Geometric Meditation

Farshad's Geometric meditation (FGM) involves six main categories of techniques. Having consulted a clinical neuroscientist as well as the originator of FGM, our treatment employed some FGM techniques from the 4 main FGM categories:

A) Mindfulness-based meditation which comprises two sub-groups of techniques: Detached Mindfulness (meditative metaphors) and Pendular Movement of Attention (PMA). These techniques train people to behave in a passive manner and without using rational processes (such as deduction, induction, analysis, conclusion, judgment, evaluation, inference or interpretation) to devolve from the intrusive thoughts, destructive emotions and around events (22). FG Metaphors (e.g., Passing clouds, flying eagle, running train, waterfall and river) can learn people the skills and strategies of how to deal with thoughts and helps someone to be able to discern the true self from defective thought processes. PMA technique helps person to not only be aware of the situation and the dependency of attention on the present moment (in a passive mode), but he can track the physical, mental and emotional problems which devour his attention like a black hole (23).

These techniques were selected for the purpose of training the client to observe her mental events (regardless of whether pleasant or unpleasant thoughts) without engaging with them, also training the client on how to regulate her disturbing thoughts, urges or impulses. For example the thought-action fusion (TAF) is known as an effect involved in OCD which has two beliefs i.e. thinking about an unwanted event increases the likelihood of its happening and the belief that an improper thought is equivalent to doing it as immoral behavior (43); and mindfulness approach could help these patients to cope with such fusions.

B) Somatic Based Meditation which include two techniques: Uni-lateral Movement of Attention (right /left) and Jumping Movement of Attention

(joint to joint) (22). The major theme in these techniques is focus on the body. They have two objectives: increment of body awareness and deep down relaxation.

These techniques were also selected to improve sensual-bodily awareness, emotional-mental awareness and also improve sleep quality (22).

C) Geometric Introspective Based Meditation comprising 10 techniques. It seems that these techniques can throw attention to the prefrontal cortex by increasing the depth of meditation (23). Three-dimensional expansion of attention technique was selected of this group to the treatment plan aims to increase the patient's attention span and scope.

D) Mind-Body Based Meditation with its efficient module known as the Figure-Background technique. According to Nazaraghaei, this technique is a combination of concentrative meditation and mindfulness-based meditation in FGM. This technique was chosen to improve both attachment and detachment skills to adjust their relationship with thoughts and other mental processes as well as the surrounding events in daily life.

### *4.2. Adapting to GM-CBT*

This treatment was delivered by the first author, who was professionally trained in coaching courses for FGM. The GM-CBT was delivered over eight 90 min-sessions, once a week. This treatment is a combination of FGM and elements of CBT associated with ERP in the form of an integrated treatment. The goals of the treatment were: Increasing insight in the patient; increasing the motivation to engage in treatment; teaching the role of maladaptive and automatic thoughts in the behavior and cycle of personal experiences of the patient, training new Skills to exposure and coping with intrusive thoughts; and reducing dysfunctional behaviors.

After familiarizing the client with the treatment model at the first session, the overall content of the subsequent sessions was:

1. Determine the agenda; 2. Review the previous homework; 3. Training based on agenda; 4. Provide new homework; 5. Review key points, provide and get feedback, and end the session.The content of the therapy sessions are outlined in Table 1.

A 45-minute meeting was held to examine the patient's progress at the end of each month. Besides, YBOCS was used again in addition to a self-report at the end of the third month.

Table1. The FGM-Based CBT protocol

| Sessions | Course content |
|---|---|
| 1 | **Psychoeducation:** Case conceptualization; Normalization; Choosing among available treatment options; Describe GM-CBT (structure & components); and Preparing the client for treatment. |
| 2 | **Psychoeducation:** Finding maintenance factors: overt and covert safety behaviors; dominant cognitive distortions & meta-cognitive beliefs (e.g. fusion of thought / action)<br>**FGM:** flying eagle metaphor. |
| 3 | **Psychoeducation:** Emphasizing the role of maintenance factors (using cycle of her personal experiences).<br>**FGM:** Free association technique in the FGM style (Understanding the difference between attachment and detachment); Pendular Movement of Attention training. |
| 4 | **Psychoeducation:** Again, the role of maintenance factors (using cycle of her recent experiences).<br>**FGM:** fish metaphor / Uni-lateral Movement of Attention training (left). |
| 5 | **Psychoeducation:** The rationale for using Exposure in 2 ways (ERP and EMR) and prepare client for them<br>**FGM:** Free association technique in the FGM style; Jumping movement of attention. |
| 6 | **Psychoeducation:** Invoke and exposure to obsessive thoughts and impulses and mindfulness-based responses (EMR); Emphasize on using ERP in real life (situational exposure)<br>**FGM:** monkey metaphor; 3 Dimensional expansion of attention training. |
| 7 | **Psychoeducation:** Emphasizing Cognitive restructuring.<br>**FGM:** waterfall and river metaphor; Uni-lateral Movement of Attention training (right). |
| 8 | **Psychoeducation:** Preparing the client for the end of treatment; Relapse prevention.<br>**FGM:** Figure-Background Technique. |

## 5. Results
### 5.1. Assessment

The total score of the fourth scale of MMPI-2 was changed from 77 at baseline to 58 after the intervention. Also, MMPI-2 Content Scales showed changes in anxiety (T=54), obsession (T=56), depression (T=52), and anger (T=60) scores. The YBOCS score decreased from 25 to 12 after eight 90-min sessions; changing from 12 to 4 after 3 months follow up.

Likewise, the total score of OBQ-44 increased from -74 to +34 and separately, changes in the five factors from pre-intervention to post-intervention included: general score change from -20 to 16; perfectionism/ certainty score change from -19 to -8; responsibility and threat estimation score change from -12 to 5; importance and control of thoughts score change from -18 to 15; and complete performance score change from -5 to 6. In addition, the total score of mindfulness was increased from 73 to 113 based on Persian version of FFMQ.

The score gains for each factor were as follow: Description of inner experience from 16 to 20 (Max=30), Concentration and mindfulness from 11 to 19 (Max= 25), Non-judging the inner experience from 13 to 21 (Max=30), Observing thoughts, emotions and physical senses from 12 to 20 (Max= 25); Acting with awareness from 13 to 20 (Max=30); and Adjust reactions to events from 8 to 11 (Max=15).

### 5.2. Close-open ended questions (for the last sessions)

a) *"Yeah...it was really effective for me. I think the entire content of the therapy sessions was good, and I think perhaps the meditation practice was more effective to me than the other."*

b) *"This therapy helped me not to feel anxious and guilty, despite unwanted sexual thoughts or impulses. In fact, they are not annoying me as I do not run away from them any longer. I am here and they are also present; and now I can concur. Since we did invoke stressful thoughts in the sixth session, they more often come up in my mind during the day, but in a very short time (less than 5 minutes), and I find it interesting that they are not bothering me, because I did not try to get them out of my mind. I just observe them. However, I still sometimes think that others can understand what's going on in my mind, but logically, I know it is a cognitive mistake and thoughts are not always equal to reality. So, I'll probably need more practice techniques. Also, I* *felt contractions, cramps and pains in different parts of my body during the first sessions; but they shortly disappeared. In particular, I suffered contractions of muscles in my forehead and around my eyes, but now the pressure is gone."*
c) *"Yeah … I decided to spend about 20-30 minutes a day to practice these techniques; because they have helped me to feel a lot better."*
d) *"I love almost all the techniques and practice them consistently. I familiarized with these techniques gradually; and now my experiences are different from my early experiences. Now, I am more deeply in 3-dimensional expansion of attention, pendular movement of attention and uni-lateral techniques. Oh… I love metaphors too."*
e) *"Yeah, I feel more relaxed when I practice these techniques. I feel I judge less about myself and others. I suppose I am kinder to myself and to others."*
f) *"Frankly, I want meditation help me with my anger management. I still lose my temper easily and talk like uninterruptedly."*

### 5.3. Self-reported experience (upon follow up)

### 5.3.1. First Month

*"I feel that I am equipped with the techniques that I've been trained in this course of treatment. Well… I have a tool box and I'm not alone! I am practicing at least one of the FGM techniques on a daily basis, and sometimes, if I have enough time, would take more than one. Muscle contractions disappeared in my body during meditation. By time, I got new meditative experiences through regular training techniques.*
*Feelings of deep relaxation, sense of expansion, sense of rotation, sense of floating, and deep-dive or vibration are some of them. I am using the ERP technique in order to expose myself to situations that I had avoided in the past, and I use the Mindfulness techniques to deal with impulsive and intrusive thoughts that I had suppressd or run away from. To tell the truth, sometimes the intrusive thoughts and the impulses invade further into my mind, but it is very strange that I do not have feelings of anxiety or guilt. Even my annoying gaze removed at the time of intrusive thoughts. Fortunately, most of the time I can I identify people could not understand what goes*

*through my mind so contraction of the muscles around my eyes have disappeared. Also, I do not judge myself, others and my thoughts like before. I'm not too strict with my subordinates, as in the past. I feel I have more flexibility.*
*Recently, my family and colleagues have noticed such changes in me. I'm so happy. Also, practicing the jumping movement of attention technique before bed has helped me to improve the quality of sleep. I now sleep well, and I do not see disturbed dreams like before."*

### 5.3.2. Second Month

*"I feel that I have become more expertise in the FGM techniques. I can use metaphors pretty much anywhere, even with open eyes, for example, at work, at parties and in the car. The thoughts and impulses still sometimes come to my mind but they did not bother me significantly. I'm observing them without having to get involved with them. It is surprising and I feel that maybe I'm too shameless and insolent (she laughs). Also, I've done jumping movement of attention technique every night during the last month, and this technique really improved the quality of my sleep. Interestingly, I do not feel tired when I wake up in the morning. Overall, I can say that I'm more relaxed and happier now."*

### 5.3.3. Third Month

*"I can say I really feel good. My mind is no longer involved intrusive thoughts or urges as previously. Maybe, sometimes they automatically come to my mind but soon disappear; perhaps because I do not attempt to drive them out of my head. The only thing that I do is that I'm observing with mindfulness. Honestly, I have not done my exercises on a regular basis this month like the past two months; but it is amazing that I am involuntarily using effects of them in decision-making, and problem solving every day. I'm living those insights. It's interesting to me how I can do it! For example, I recently had a lecture, but without such feeling of anxiety for the presentation. I feel I am more confident than before. Also, I can do anger management better than before. Meditation has even affected my diet; I do not overeat as in the past. I was living in the dark for many years until I entered this. I trust these meditation techniques were the breakthrough in my life. It is incredible how I left behind all those sufferings with the help of this treatment."*

## 6. Discussion

According to the present case investigation, the FGM-based CBT intervention was quite effective in ameliorating the client's OCD symptoms. Based on our earlier research, pre- to post-intervention decrease of ≥ 35% on the YBOCS indicates a proper response to therapy and a score of ≤ 12 is the best prophesier of wellness (44). So, a 13-point reduction in the Yale-Brown scale indicates a clinically meaningful change in present study after 8 sessions. Also, the obsessive beliefs improve 108 score based on the OBQ-44. In addition, her mindfulness scores increased 40 scores after treatment completion. Besides, the total score of the fourth scale of MMPI-2 and MMPI-2 Content Scales scores (anxiety, obsession, depression, and anger) were decreased (fell in normal range).

Moreover, qualitative report on the status of the patient (self-report) shows the optimal situation for her. In fact, the primary outcomes of this case report show a significant decrease in OCD symptoms and beliefs as well as anxiety and depression symptoms. Also, an increase in mindfulness skills was observed. According to client self- reports, there was an increase in insight, loss of musclel contractions and improved in sleep quality. Interestingly, an additional 9-point reduction in the YBOCS was observed at the end of 3 months follow-up. Other self-reported secondary outcomes of the treatment were improvement in the anger management, overeating, self-esteem, performance anxiety, and flexibility in interpersonal relationships.

Based on the behavioral theory, negative reinforcement is known as a perpetuating factor for compulsive behaviors in OCD (45). So, ERP was developed as a behavioral therapy (or as a component of CBT) for OCD and other anxiety disorders (8). Evidence shows that about 25% of OCD patients abandon ERP therapy; and about half of them fail this treatment probably due to anxiety-provoking essence of ERP (46-47). This is perhaps because patients are not well-equipped with coping strategies to deal with anxiety and stress, hence it is not surprising if they fail to tolerate such treatments.

Today, the effects of acceptance and mindfulness were approved as two important components in the third wave of cognitive-behavioral therapies. Acceptance, non-judgmental observation, and regulating negative emotions which are mindfulness components may help abandon someone from dysfunctional and maladaptive responses such as avoidance, repression and obsession-compulsion (48). In mindfulness-based cognitive therapy (MBCT), unlike ERP, clients are trained to observe inconvenient experiences and accept them as thoughts, emotions or bodily sensations; but unwanted experiences are not summoned to the mind, consciously (49). Therefore, one may use mindfulness techniques as distraction techniques, or a way to neutralize or suppress intrusive thoughts, unconsciously.

In our treatment model though, CBT was used for the purpose of case conceptualization as well as psycho-education. And FGM techniques were used to materialize mental contents as well as the skill of mindfulness, cognitive detachment, the expansion of awareness, confront bodily sensations and muscle relaxation. We have used FG Meditation techniques before the exposure technique in this treatment protocol. The objectives were three-fold: a) client involved in the treatment, without focusing on her anxiety-provoking issues; b) the individual could equipped with coping skills helping her to cope with the anxiety, worry and thought ruminations which intruded upon the healing process; c) the applied techniques could help the client how to deal with her mental processes through a whole new adaptive strategy.

Furthermore, obsessive thoughts and impulses of the client in the sixth session recalled deliberately and asked her to be confronted with them by mindfulness skills instead of suppress, neutralize or avoid the obsessive thoughts (her past coping); what we named EMR that means voluntary Exposure with intrusive thoughts and Mindfulness Response. The aim was to ensure the client does not use the techniques of mindfulness as a distraction technique, or as a way to avoid or escape them unconsciously. Because, the ultimate goal of this therapy was to instruct the client to observe the unpleasant contents in her mind as a "passive observer" rather than engaging with them or using dysfunctional strategies such as neutralization, prevention, or suppression.

Given these cases, the use of EMR and ERP (situational) makes the individual to understand that anxiety, fear, uncertainty, obsessions, and compulsions are manageable; so there's no need to avoid or escape them. They may decline over the time in the absence of triggers in real life or through repeated confrontation; also, bodily sensations are not cues for threats, and contractions and cramps in muscles gradually will disappear after persistent exposure.

We trust that this case investigation would only be the beginning of the research path towards the application of FGM in mental illnesses. Apparently, evidence-based data would need to compile before one can advocate the effect of FGM in ameliorating complex symptoms of OCD. To this end, further research using a prospective randomized controlled set up would be warranted.

## *7. Conclusion*

Our findings suggested that GM-CBT for OCD symptoms may exert beneficial treatment effect thus far at a case study level of evidence.